\documentclass[letterpaper]{article} 
\usepackage[]{aaai25}  
\usepackage{times}  
\usepackage{helvet}  
\usepackage{courier}  
\usepackage[hyphens]{url}  
\usepackage{graphicx} 
\usepackage{natbib}  
\usepackage{caption} 
\usepackage{algorithm}
\usepackage{algorithmic}

\usepackage{newfloat}
\usepackage{listings}
\DeclareCaptionStyle{ruled}{labelfont=normalfont,labelsep=colon,strut=off} 
\floatstyle{ruled}
\newfloat{listing}{tb}{lst}{}
\floatname{listing}{Listing}
\usepackage{todonotes}

\title{A Case for Specialisation in Non-Human Entities}
\author {
    El-Mahdi El-Mhamdi\textsuperscript{\rm 1},
    Lê-Nguyên Hoang\textsuperscript{\rm 2},
    Mariame Tighanimine\textsuperscript{\rm 3, \rm 4}
}
\affiliations {
    \textsuperscript{\rm 1}École Polytechnique, France\\
    \textsuperscript{\rm 2}Calicarpa, Switzerland\\
    \textsuperscript{\rm 3}Lise, Cnam CNRS, France\\
    \textsuperscript{\rm 4}University of Neuchâtel, Switzerland\\
    mariame@tighanimine.com\footnote{Corresponding author}
}

\usepackage{bibentry}

\begin{document}

\maketitle

\begin{abstract}
    With the rise of large multi-modal AI models, 
    fuelled by recent interest in large language models (LLMs), 
    the notion of artificial general intelligence (AGI) went from being restricted to a fringe community, 
    to dominate mainstream large AI development programs. 
    In contrast, in this paper, we make a \emph{case for specialisation}, 
    by reviewing the pitfalls of generality and stressing the industrial value of specialised 
    systems.
    
    Our contribution is threefold. First, we review the most widely accepted arguments \emph{against} specialisation and discuss how their relevance in the context of human labour is actually an argument \emph{for} specialisation in the case of non human agents, be they algorithms or human organisations. Second, we propose four arguments \emph{in favor of} specialisation, ranging from machine learning robustness, to computer security, social sciences and cultural evolution.  
    Third, we finally make a case for \emph{specification}, discuss how the machine learning approach to AI has so far failed to catch up with good practices from safety-engineering and formal verification of software, and discuss how some emerging good practices in machine learning help reduce this gap.
    In particular, we justify the need for \emph{specified governance} for hard-to-specify systems.
\end{abstract}

\section{Introduction}

In 2023, the European Union announced its Artificial Intelligence Act. The AI Act consists in a series of regulations on AI that not only impacts European AI, but ultimately also other countries and regions in the world in what is known as the Brussels' effect~\cite{bradford2020brussels, bradford2024brussels, siegmann2022brussels, gunst2021brussels}. 
One of the most puzzling features of the AI Act is that, due to intensive lobbying from large AI companies~\cite{perrigo2023exclusive}, the Act ended up putting more restrictions on some specialised AI models than it does on AI models claiming general capabilities. To capture the absurdity, consider an analogy: in medical practice, would a general practitioner have more freedom to operate on a patient's eye than an ophthalmologist? The answer is obviously negative. The less specialised a medical doctor, the more restrictions different medical legislation or codes of ethics\footnote{See e.g., ~\cite{american1871code, codesante}.} would impose on what they can do.

In the 2020s, AGI, in which the ``G'' stands for \emph{general}, has increasingly become an overused term to include several \emph{desirable} properties in AI entities.
However, it is unclear whether \emph{generality} ought to be regarded as \emph{desirable},
especially in terms of auditability, reusability and security, 
but also in terms of industrial value.
In this paper, we review social science and statistical arguments \emph{for} generality, 
and highlight their limits in the context of non-human entities.
We then instead make a case \emph{against} generality, 
by leveraging arguments from adversarial machine learning, 
from complex system engineering 
and from social sciences. 
We then stress the value of \emph{specification}, which requires specialised systems.

The rest of this paper is organised as follows. In Section~\ref{sec:context}, we lay down the broader context of our work and review useful notions such as \emph{generality}, \emph{task} and what we call ``\emph{friends of specialisation}'' such as decentralisation or separation of powers. In Section~\ref{sec:against}, we propose three arguments against specialisation from social sciences, economics and statistics, and discuss their limitations in the context of non-human entities. 
This brings us to Section~\ref{sec:prospecialisation}, in which we propose our arguments for making non-human entities specialised. 
Section~\ref{sec:specification} complements our argument by making a case for \emph{specification}.
The section also discusses the limits of specification, 
and the need of \emph{specified governance} to address these limits.
Finally, \ref{sec:conclusion} concludes this paper.

\section{Context}
\label{sec:context}

Historically, the word ``algorithm" comes from the name of Muhammad ibn Musa al-Khwarizmi, 
and really consisted of decomposing complex tasks into elementary trivial subtasks, just like in al-Khawarizmi's pioneering book, providing simple step-by-step recipes to simplify problem resolution for a wider audience, and in particular, inheritance case resolution for lawyers.

This intuitive approach, insisting on breaking problem resolution to simple tasks, was further formalized by the pioneers of computer science, 
in particular Alonzo Church~\cite{church1936note} and Alan Turing~\cite{turing1936computable}.
Both argued that any computable function can be written as a composition of elementary ultra-specialised hard-wired mechanical operations,
typically on a Turing machine (or in lambda calculus).
In today's computing machines, 
such operations are the (now numerous) logical circuits engraved in processing units like CPUs.
The assertion that these fixed circuits suffice to perform 
\emph{any} computable information processing function is now known as the Church-Turing thesis.
This thesis arguably thus asserts that, in principle, 
any general capability is a composition of (a very large number of) specialised operations.

However, in practice, when software consumers use an algorithm,
the compositions of such specialised operations is abstracted away,
so that the consumer ends up interacting with a complex system,
whose range of capabilities can be narrow or wide.
Lately, the rise of language models, but also that of fully integrated and connected cloud services like Microsoft 365, has led to the commercialisation of ``general systems". 
Such systems (are said to) provide a large number of information processing services.

In fact, many software developers might not view their software solutions 
as a composition of specialised operations, especially if their solutions leverage external libraries. 
While some of these codes simply translate the developers' programming language codes into elementary binary codes (these are \emph{compilers}), it has become extremely common for developers to use complex ``general-purpose" libraries, especially when importing, e.g., large language models. 
Such systems could be argued to ``empower" developers, 
as they can now effortlessly program solutions to many more problems.
However, they can also be argued to reduce their capability to understand their own systems,
especially in terms of capability, safe usage and emergent risks.
In contrast, developers could favor building upon \emph{specialised} libraries,
which provide them with precisely the tools they needed for some well-defined tasks,
such as optimization solvers.
The developer would then only be blind to such specialised operations,
which potentially having the capability to clean and sanitize their outputs.
It is noteworthy that such specialised libraries could be made to be equally blind to how they are used.

What we might call ``specialisation blindness'' can also be found in other activities or professions. A fast-food franchisee, for example, considers that their business is selling hamburgers or other products sold in this type of restaurant. However, upstream, a franchisor has to call on several specialist trades to develop all the building blocks for a successful franchise: the right furniture (interior designers, furniture salesmen...), the right recipes (chefs, nutritionists, etc.), the right ingredients (farms, bakers, drinks wholesalers...), the right business model (bankers, accountants...), the right location (estate agents, notaries...), etc.

Specialisation blindness is arguably an unavoidable consequence of collaborating to perform complex tasks~\cite{boullier2020machine, tighanimine2025travail}.
However, the question we raise in this paper, is how to organise collaboration to make it an asset
rather than a risk.
Essentially, we will argue that the key lies in well-specified specialisation.

\subsection{Defining Generality}

In the context of the AI Act, generality is often associated with the \emph{purpose} of a system. The greater the diversity of use cases, the more ``general-purpose" the system. This definition of generality-by-purpose can be also found in Model Cards~\cite{mitchell2019model}, which asks AI system developers to specify the ``intended use cases" of the developed systems.

We note however that there may be ambiguity behind the definition of a use case.
Assume for instance that a language model is used to translate messages on a social media platform.
Is the use case ``social media"? 
Or merely ``translation"?
Or even more simply ``next-token predictions", given an original text and a translation request prompt?
Clearly, if we consider the first answer (``social media"),
then the system may appear very ``general-purpose";
but more basic algorithms like json serializers or symmetric encryption should then be regarded as more ``general purpose".
Conversely, the more we dig into the very precise use of a system (``next-token prediction"), the less ``general-purpose" it will appear to be.

A leaked document~\cite{Maxwell2024leaked} from OpenAI and Microsoft revealed their use of a more financial approach to defining AGI.
Namely, they (privately) regarded as ``general" a system that generates hundreds of billions of dollars in revenues.
In a sense, this approach is similar to counting use cases of a system,
but it proposes to weigh the use cases by their financial added values.
It is however noteworthy that, given this definition, 
Blackrock's trading algorithm Aladdin \footnote{In 2020, Aladdin managed 21.6 trillion U.S. dollars in assets \cite{Ungarino}.} and 
Google's ad targeting algorithm AdSense \footnote{Google's ad revenue amounted to 237.86 billion U.S. dollars in 2023 \cite{Statista}.} should be regarded as much more ``general" than language models.

Another approach that may be considered to assess the generality of a software system
is the list of \emph{application program interface} (API) calls it defines.
Indeed, each API function proposed by the system may be viewed as a task that the system can offer to solve.
Of course, in practice, some API functions can be regarded as themselves more ``general" than others,
e.g. if they involve multi-modal inputs rather than text-only.
While still not fully rigorous, in the sequel,
we lean towards this definition.
We will regard a system as general, 
if the number of tasks that external users can ask the system to perform is large.

Note that a very specialised system in this sense may nevertheless be extremely complex.
For instance, even though a language model is only accessible through a ``next-token prediction" API,
or if its only API yields content recommendations,
it may be itself composed of trillions of parameters \cite{DBLP:conf/kdd/LianYZWHWSLLDLL22}.
In fact, any large human organization that is specialised to deliver a very specific task
must perform a large number of internal tasks,
such as accounting, sales or human resource management.
Nevertheless, we will regard it as specialised if the number of services 
it offers to \emph{external users} is small (and well-specified).

\subsection{AI Agents}

Since 2024, most large AI companies have been promoting AIs 
with \emph{acting} capabilities~\cite{yao2023react,acharya2025agentic,murugesan2025rise,borghoff2025human}.
As opposed to conversational AIs that merely send messages to end users,
such so-called \emph{AI agents} (or \emph{agentic AI}) can execute commands on information systems.
More precisely, the Model Context Protocol (MCP)~\cite{hou2025model,ray2025survey}
was recently proposed to standardise the way AI agents do so.
Namely, each AI agent is given a list of API calls that they can call,
along with the documentation of these API end points.

It is noteworthy that such agents differ 
from other concepts of \emph{autonomous agents} 
that rather refer to the capability of such agents to behave well
in some general frameworks, like reinforcement learning~\cite{hutter2003gentle,veness2010reinforcement},
often with respect to a given objective function that sends rewards~\cite{DBLP:conf/aaai/Ben-PoratMMT24,DBLP:conf/aaai/KieransGHD25}.
Such frameworks typically do not invoke the access to API calls.

In some regards, MCP thereby somewhat formalizes the concept of generality we discussed earlier.
In particular, one could be tempted to give to an AI agent 
a large number of API call accesses,
to enable them to solve complex tasks.
Our case is especially \emph{against} such developments.
In particular, like others~\cite{blili2025stop,radosevich2025mcp}, 
we argue that it is not in the interest of companies and societies
to integrate such agents in their organizations.

\subsection{Defining the Notion of Task and its Granularity}

Providing a single definition of what a task is can be tedious. We can start by trying to understand its centrality in automation and in the more general context of work (which inspires our thinking on specialisation), in division of labour.

In a work context, a task can be considered as an action associated with goals, means and conditions of execution. Within the context of prescribed work, a task corresponds to all the goals and procedures defined in advance, and meets codes, performance requirements, and quality standards. The elements of prescription - even though they may be \emph{underdeveloped} - are generally found in all work activities. Various institutions, professional hierarchies, public authorities or professional groups set tasks, objectives, procedures, directives, rules, and decrees defining what can be done or must be done.
Social sciences of labour have repeatedly shown by means of empirical studies that what workers, managers and even executives do is not simply the execution or application of the task prescription. Everything they do systematically goes beyond the divide between ``task'' (what is prescribed) and ``action'' (what is done). In the language of computer science, these may be translated as ``specification" and ``code execution".

When we look at the history of the organisation and rationalisation of work, particularly through the emblematic examples of Taylorism \cite{littler1978understanding}, Fordism \cite{watson2019fordism} and Toyotism \cite{dohse1985fordism}, all of these paradigms have involved defining and sorting out specific tasks, selecting workers to carry them out, and more generally (over)specialising work and then automating it. For the past two centuries, automation has been the hallmark of work transformation. Broadly speaking, it starts with an agent's craft know-how. Automation then consists of extracting the skills from the agent, and of encoding them into a procedure,
i.e. a series of simple, structured and repetitive tasks be they physical or cognitive.
In debates about the ``future of work''~\cite{brynjolfsson2017can}, jobs that are identified as least likely to be automated are those whose tasks are unstructured, non-routine, and whose performance is associated with critical thinking, long chains of reasoning or complex planning, a certain level of creativity, etc.
Arguably, this is because the resolutions of such tasks are hard to encode into procedures, though the rise of machine learning opens the door to the encoding of procedures that are humanly hard to describe, e.g. in the Kolomogorov-Solomonoff sense (see Section~\ref{sec:solomonoff}).

Automation - the creation of machines and algorithms to carry out tasks previously performed by humans - is a constant reconsideration of division of labour. To quote Karl Marx (himself echoing Charles Babbage), machines emerge as a synthesis of the division of labour~\cite{marx2019misere}. So the notion of task is at the heart of automation. An important question is to know what level of definition and operations to adopt in order to automate tasks and, ultimately, specialisation. There are limitations that make this exercise complicated. 
A prominent challenge is to select the granularity of a task (and thus its level of specialisation). There are informational and computational hurdles that make this exercise complicated. For example, there are many tasks that we understand tacitly and carry out without being able to state their explicit rules and procedures. In other words, formulated by Polanyi's paradox \cite{autor2014polanyi}, ``we can know more than we can tell’'~\cite{polanyi2009tacit}.
Moreover, automating certain tasks and not others also means renouncing possible ways of working and other possible uses~\cite{simondon1958mode}.

Defining a task can also refer to expressions such as granularity or modularity. We can define granularity and modularity using the definitions proposed by Benkler \cite{Benkler} which are the basis of his socio-economic production model ``Commons-based peer production''. Modularity is a way of dividing a project into smaller components (modules) that can be developed independently before being assembled to form a whole, in order to maximise the flexibility and autonomy of the contributors. Granularity refers to the size of the modules, defined in terms of the time and energy required by the participants to produce them. It is important to specify the modules and their quantity in a project, because the interest and investment of the participants depend on it. As we will see, the two concepts are also central in computer science and system engineering.

\section{The Case Against Specialisation}
\label{sec:against}

Before making our case for specialisation, 
let us highlight the most common arguments against specialisation, 
and discuss why they feel ill-fit to today's AI systems.

\subsection{The Case of Humans}

\paragraph{Overspecialisation harms workers' mental health and flourishing.}

Although some arguments may have been anterior, the most severe criticisms of the division of labor opposed Taylorism, assembly line work, and what came to be known as Scientific Management\cite{bookwalter1918scientific}. This work organisation generated an extreme division of labour with a fragmentation of tasks, a sustained pace of execution, a high degree of dependence between workers, and the impossibility of envisaging the unexpected events. Criticisms from the social sciences of work were to emerge in the middle of the twentieth century and the rise of increasing industrialisation (particularly in the automobile sector) and mechanisation. Friedmann \cite{friedmann1955travail, friedman1961anatomy} talks of ``work in crumbs''\footnote{Better captured by the original French expression ``Travail en miettes''.}” to describe the evolution of forms of work with a high level of automation, and in particular the work in the factories. There have also been concerned with the psychological damage caused to workers by alienating, unskilled, and depersonalized work.

Today, criticisms of the excessive division of labor and hyper-specialisation concern many economic sectors (food industry, logistics, textile industry, etc.). They particularly focus on the aspects that are detrimental to the body and mind of individuals. One of these industries is precisely the information and communications technology industry, and particularly digital platforms (on-demand platforms, social media), and more generally artificial intelligence using digital labor, micro-work and data workers, especially in the global south countries~\cite{casilli2020virtual, williams2022exploited}.

\paragraph{Overspecialisation leads to a loss of meaning/purpose.}

High specialisation can cause workers to lose interest and motivation in their work, and can impact performance \cite{loukidou2009boredom}. A high level of specialisation tends to lower workers’ stimulation and motivation levels, while increasing their boredom and disengagement \cite{hackman1969nature, mccauley1994assessing}. 

However, while the psychological and physical damages on human workers are important to consider, algorithms are arguably not subject to such concerns. At least following~\cite{gibert2022search}, while they could be regarded as \emph{moral agents} (i.e. they have moral duties), they should not be regarded as \emph{moral patients} (i.e. there is no moral duty to protect them). Therefore, the above arguments do not seem to apply to non-human agents.

\subsection{Arguments from Economics}

In economics, instead of specialising in a given step of the production chain, 
companies may be tempted to invest in \emph{vertical integration}~\cite{coase1937,bresnahan2012vertical}. 
Namely, they could want to own their entire supply and delivery chain. 
The theoretical analysis of \cite{arrow1975vertical} indeed found that imperfect information can incentivize such vertical integration, 
while that of \cite{carlton1979vertical} instead stresses the role of transaction costs.
\cite{grossman1986costs} further underlined contractualisation costs:
if specifying the conditions of an agreement between a supplier and a producer is challenging,
then there will be strong incentives for vertical integration.
The theories of \cite{arrow1975vertical} and \cite{carlton1979vertical} have both found support,
e.g. by \cite{lieberman1991determinants} in the chemical industry.
On top of this, \cite{fetz2010economies} empirically observed
better investment coordination and less financial risk 
following the vertical integration of the Swiss electricity sector.
In practice, varying working conditions may instead incentivize outsourcing,
and thus vertical disintegration~\cite{ricardo1817,kakabadse2005outsourcing}.

In the software industry, 
yet another phenomenon may incentivize both vertical and lateral integration,
namely the \emph{network effect} \cite{shapiro1999information}.
This describes situations where the value of a product is increased by the wide use of this product,
or of similar related products.
A classical instance of this is Facebook's huge investments in social media.
For instance, by acquiring Instagram and by connecting it with their other platforms,
Facebook has exploited the network effect to increase the value of their anterior assets~\cite{li2017platform}.
Similarly, Google and Apple now produce both hardware, operating systems and applications,
thereby offering a unifying solution to their customers.
The value of the hardware is then augmented by a corresponding optimized optimized system,
which is itself augmented by its connection to a dedicated cloud system \cite{vergara2012samsung}.
Further, software companies have incentives to invest in lateral integration,
to propose all-in-one highly connected digital workspaces,
as is the case for instance of the Microsoft 365 product.
This further helped them push AI-based solutions, such as Microsoft co-pilot \cite{skendzic2012microsoft}.

However, while there may be some positive consequences for some users,
critics point out that such integration incur societal risks, 
especially in terms of market power \cite{landes1997market}.
In particular, the technology sector has been repeatedly criticized
for its dependence on a small number of actors,
as exemplified by the landmark antitrust case against Microsoft~\cite{economides2001microsoft} and the current ongoing cases against Google and Facebook~\cite{brennan2025us},
and by the numerous recent calls to enforce antitrust laws
to other actors~\cite{munir2024google,davies2024google,worsdorfer2024apple}.

Additionally, integration can increase systemic risks. 
This is well illustrated by the report of the Cybersecurity and Infrastructure Security Agency on the 2023 Microsoft Online Exchange Incident \cite{cisa}. 
The penetration of Microsoft's cloud infrastructure by foreign actors has not only compromised the targeted US institutions; it may also be endangering \emph{all} of Microsoft's customers. 
Moreover, the high connectivity of all of Microsoft's cloud services means that a given customer may be compromised, even if they only use some of these services.

Having said this, at the scale of a country,
especially in a context of tensed geopolitical tension,
vertical integration could in fact be a consideration of national security. 
More generally, it could be in a given system's best interest to seek generality, 
if they want autonomy and resilience.
However, it may not be other others' best interest to depend on a general system, which is the consideration we will focus on.
Indeed, we recall that our case \emph{for} specialisation is focused on \emph{external uses} of a system.

\subsection{Arguments from Statistics}

In artificial intelligence, 
much of the recent interest for general AIs derives 
from the observed ``scaling laws"~\cite{kaplan2020scaling},
which argue that learning from all sorts of (non-specialised) data 
is the driving force of dramatic performances.
While these observations have been criticized~\cite{diaz2024scaling},
they do have some theoretical backgrounds.

In particular, they may be argued to be 
an instance of Stein's paradox~\cite{efron1977stein}.
Strikingly, this paradox shows that, 
when performing statistical learning on three or more disjoint subsets of data, 
it is statistically \emph{inadmissible} to learn separately on each subset.
More precisely, for each subset, 
consider any (specialised) estimator of a ground truth statistics of the subset.
Then there is a (general) estimator that learns from the union of all subsets such that,
no matter what the ground truth is,
the general estimator's expected mean square error will be lower than the specialised estimators'.
Moreover, for at least one value of the ground truth, it is strictly lower.
The general estimator is said to strictly dominate the specialised estimators.

This is especially evident in the case of collaborative filtering~\cite{ekstrand2011collaborative}.
To better optimize what content should be recommended to a given user,
it is extremely useful to leverage the preferences of similar users,
rather than to learn exclusively from the given user's data.

While this statistical argument is very compelling,
it is important to highlight two key weaknesses.
First, it is a statistical, and thus \emph{information-theoretical}, argument.
It thus neglects the computational costs.
In particular, the general estimator may require significantly more resources than the specialised estimators.
And while we use a very computer-science terminology,
this remark holds for both algorithms and human organisations.
One striking example is that of science:
scientists have self-organised themselves in specialised communities.

But more importantly, the subsets of data have thus far been assumed to safe to learn from.
However, in practice, most datasets raise privacy and poisoning issues.
On one hand, general estimator can more easily cross information 
that would allow for user de-anonymisation,
even if ``privacy-preserving learning" solutions like differential privacy are used \cite{DBLP:conf/sigmod/KiferM11,DBLP:journals/tifs/ZhuXLZ15}.
On the other hand, malicious users can more easily poison the union of all subsets,
thereby biasing or harming the general estimator~\cite{DBLP:conf/icml/BiggioNL12,DBLP:conf/icml/SuyaMS0021,DBLP:conf/icml/FarhadkhaniGHV22}.
In particular, the introduction of collaborative filtering has given fake accounts
an enormous influence on the daily information exposure of billions of humans.

\section{Why Specialise?}
\label{sec:prospecialisation}

Now that we have highlighted the limits of the arguments \emph{against} specialisation,
we turn our attention to our case \emph{for} specialisation.

\subsection{Large Models are More Vulnerable}

In the context of artificial intelligence,
generality is obtained by training large models on massive web data crawls.
Indeed, ever larger models and datasets have been empirically observed to
yield ever more spectacular performances~\cite{DBLP:conf/nips/BrownMRSKDNSSAA20,DBLP:journals/corr/abs-2001-08361,DBLP:journals/corr/abs-2407-21783}.
However, these high performances seem inevitably entangled with a number of security issues~\cite{el2022impossible,oprea2023adversarial},
like privacy violations, jailbreaking and data poisoning.
Below we detail the reasons of this entanglement.

\subsubsection{The more parameters, the more vulnerabilities}

There is now a large literature that shows in particular that
the number of parameters of a learned model increases its vulnerability,
especially when it comes to privacy~\cite{DBLP:conf/compgeom/KattisN17} 
and poisoning~\cite{DBLP:conf/icml/MhamdiGR18,hoang2024poison, guerraoui2017robustness}.
Intuitively, this is because the leading solution for privacy,
namely \emph{differential privacy}~\cite{DBLP:conf/tcc/DworkMNS06},
requires adding a bit of noise to all parameters when updating the model,
in order to correctly blur any trace of the data that led to this updating.

Similarly, to defend the model against malicious training data injections,
the leading solutions essentially boil down to outlier removal.
However, in high dimension (i.e. high number of parameters),
random (honest) data are scattered away,
which give poisoners a lot of room to bias the training 
without appearing to be out of distribution.

In the case of jailbreaking~\cite{DBLP:conf/icml/GuoYZQ024},
a key variable is the input size of the model,
also known as the \emph{context window} for language models.
Intuitively, the larger this input size, 
the more the model can be tuned for specific applications via input injection 
(prompting in the case of language models).
But then, the greater the \emph{attack surface} for jailbreaking attacks~\cite{anil2024many}.

\subsubsection{The more generality, the more data heterogeneity, the more vulnerability}

There is an additional subtler way through which the quest of generality harms resilience to attacks.
Namely, to gain capabilities in a wide variety of use cases, 
the model needs to be trained on a highly heterogeneous set of training data.
Such data are typically obtained through web crawls~\cite{baack2024critical},
which include dubious sources~\cite{schaul2023inside}.

In particular, under such high data heterogeneity,
sensitive and malicious data are significantly harder to identify and remove.
More rigorously, many papers have drawn a clear connection
between machine learning vulnerability and data heterogeneity~\cite{el2021collaborative}.

\subsection{Arguments from Complex System Engineering}

A fundamental and ubiquitous principle of complex system engineering is \emph{abstraction} \cite{DBLP:conf/iwssd/Shaw89}.
This corresponds to hiding the complexity of the system,
and to highlight instead the key features of the system
as well as the limited number of ways through which it can be used.
Perhaps, the most important example of such an abstraction is the creation of programming languages.
Such languages allow programmers to define how they may get machines to do what they want,
without having to care about the precise ways their programs will lead to
transistor operations on the machines.

Abstraction is especially instrumental to implement 
the principles of \emph{modularisation}
and the ``separation of concerns"~\cite{parnas1972criteria,tarr1999n}.
Modularisation consists of dividing a complex systems 
into interacting \emph{specialised} components, called \emph{modules}.
By carefully defining the features of the modules,
which amounts to defining the abstraction they must comply with,
the development of each module can then be performed independently from other modules.
Similarly, each module can be evaluated, stress-tested and even correctness-proved
independently from the rest of the complex system.

\emph{Separation of concerns} consists of segmenting a computer program into several parts. Each of these parts is isolated and takes charge of a piece of concern or information from the general problem being dealt with. This practice simplifies the development and maintenance of computer programs. A good, strict separation of concerns means that different parts of the code can be reused or modified independently, or that work can be done on one part of the code without having to know the other parts.

Modularisation is a fundamental technique of computer science
that strongly highlights the value of \emph{specialisation}.
For instance, the hashing function SHA-256~\cite{penard2008secure} is highly specialised.
However, it is this amount of specialisation that has facilitated its thorough study for decades,
and that has made it a core module of virtually all modern complex software systems.

Better yet, clearly defining the features of each module of a complex system
helps identify the module's \emph{least required privileges}~\cite{saltzer1975protection}.
Thereby, optimized security constraints can be enforced to the complex system's modules.
Typically, an AI model used in inference mode should not be given access to the Internet,
nor to the file systems, the webcam or the keyboard.
It should only be given the input data, and the required computing hardware to perform its task.
Thereby, if a given module is flawed, hacked or backdoored, 
then the scale of the harm to the complex system will then be limited to the privileges that were given to the module. This safety oriented mindset resembles two social organisation principles: \emph{separation of powers} and \emph{Subsidiarity}. Separation of powers is the foundation of constitutional states and democracies \cite{locke1690}, \cite{de1989montesquieu}, and prevents the concentration of legislative, executive, and judicial powers in the hands of a single individual or political group. The separation of powers is not applied in the same way in all countries \cite{bellamy2017rule}. The stricter it is, the more distinct, specialised, and separate the powers are, while retaining reciprocal means of action. The more flexible, the more the different powers work together. Subsidiarity is associated with decentralisation. Inspired by the social doctrine of the Catholic Church, subsidiarity \cite{evans2014global} is a political principle that assigns responsibility for a problem to the lowest competent level of public authority. This means looking for the level that is most relevant and closest to the people affected by a decision. In this way, the higher level is only called upon if the problem to be dealt with exceeds the lower level, and what can be done with the same efficiency at a lower level must be done without a higher level.

Even with limited privileges, a module may still be very dangerous.
This is typically the case if its output is not carefully sanitized 
by the subsequent algorithmic (or human) modules.
For instance, if the output of a language model is used directly for email response,
then the language model can be exploited to spread a spam worm~\cite{DBLP:journals/corr/abs-2403-02817}.
Likewise, a recommendation algorithm can threaten democracies by merely suggesting content,
if human populations enact upon the hate speech that the algorithm amplifies.
Tricking other modules can thus yield \emph{privilege escalation}~\cite{ozdemir2024privilege}.
But crucially, by carefully specialising the modules,
it will be much easier to then reason about the risks of their usage.

Finally, the careful modularisation of a complex system can increase its availability
by implementing the interoperable redundancy of each module,
with diverse implementations of the module.
The simplest example of this is the case of storage replicas.
However, more inspiring systems with interoperable modules have been designed for complex tasks.
Recently, the AT Protocol proposes to segment the tasks of a social media,
with interoperable modules such as data storage, 
published message collection (know as ``relays"),
message labeling,
feed generators
and client-side applications (known as ``AppViews")~\cite{kleppmann2024bluesky}.
By encouraging and facilitating the creation of interoperable modules managed by other entities,
the protocol helps reduce the risks of social media network effects~\cite{dou2013engineering}.
Similarly, \cite{solidago} proposes a modularisation of collaborative scoring,
as a way to facilitate the construction of numerous interoperable subtasks.

All of these elements naturally bring to mind the notion of \emph{decentralisation}, studied in various fields (computer science, political sciences, management science, law, public administration, economics, etc.) with different connotations. For example, in social sciences and particularly political sciences \cite{treisman2007architecture}, it usually refers to delegation of power, transfer of skills and resources from a central power to authorities/local communities distinct from it.

Most of the notions we have mentioned (subsidiarity, decentralisation, separation of concerns and separation of powers) are sometimes used interchangeably with specialisation, and are involved in positive feedback loops with specialisation and while not delving into the interdependence between these notions and specialisation in this work, we highlight this noteworthy relationship that makes many of the arguments made in our paper at least partially valid for subsidiarity, separation of concerns, separation of powers and their alike, with maybe the notable exception of decentralisation. 
Decentralisation, and generally delegation of power, also offers some level of generality. For instance, components of a decentralised learning system can be learning all the same tasks, just as states in a federal system all have prerogatives on most administrative and daily life needs. Decentralisation and delegation offer advantages of their own, such as isolation of faults or confinement of corruption to preserve the overall system.

\subsection{Arguments from Economics}

\paragraph{Specialisation Increases Competitiveness.}

The notion of specialisation is often associated with \textit{division of labour}, which has even been a major theme in philosophy and economics for centuries (\cite{marx1867}). In particular, division of labour should not be reduced to its technical dimension. Namely, the interdependence it generates between individuals within a society or on the scale of several has led to considering the division of labour as a important brick in the foundation of societies. 

While the division and specialisation of labour within human societies can be dated back to the Neolithic period, they rather started being discussed in the 16th century. Their formalization in a systematic thinking is often associated with the writings of \cite{smith1776}, who explored the logic behind the fragmentation of work and the specialisation of tasks, as well as the links between division of labour and market competition. Smith used the example of a pin factory to illustrate the division of labour, with a work decomposed into eighteen operations required to produce one unit. He notes that each step of work fragmentation increases productivity. Unlike craftsmanship, where the craftsman controls the entire production process, manufacturing is superior in terms of productivity, due in particular to the simplification of tasks, the reduction of downtime, and control over the pace of work. 
He does, however, point out one limit to the division of labour: the size of the market. If the market is not big enough, there will be no outlet for the surplus production resulting from an ever-increasing division of labour.

While Smith focused on the specialisation of workers on a single production line, still within classical political economy, \cite{ricardo1817} examined the division of labour on a global scale. 
With his theory of comparative advantage, Ricardo analysed the specialisation of national economies as they opened up to trade, using the famous example of England and Portugal and their production of wine and cloth. Ricardo's theory has later been further developed in numerous works on the international division of labour and international specialisation (\cite{samuelson1948international, HeckscherOhlin}). 

Note that, while Smith focused on human agents, Ricardo and others have already stressed the value of specialisation for non-human agents,
by arguing that it increases economic competitiveness and global production. Indeed, the theory of comparative advantage asserts that,
perhaps surprisingly, a less effective agent could still be helpful to a more effective agent, if the less effective agent specializes in tasks that they perform best, even in cases where the more effective agent performs better at these tasks.

\subsection{Arguments from the Sociology of Work, Occupations, and Organisations}

\paragraph{Specialisation enables intermediate bodies.}
    
Beyond the considerations of economic competitiveness,
Durkheim put forward a thesis\cite{durkheim1893division} asserting that the division of labour is the source of the social bond in industrial societies. He even made the study of changes in the division of labour one of the foundations of sociology. The division of labour is, he says, a source of “organic solidarity” characterized by differentiation, cooperation and strong interdependencies between individuals, as opposed to mechanical forms of social organization - individuals grouped together in communities on the basis of proximity, authority or faith. Aware of working-class misery and social conflicts, Durkheim advocated in the preface to the second edition of his book (1902) for a greater application of the division of labour to social organization, through the creation of intermediate bodies constituted on a professional basis, so that they could play the role of moral authorities arbitrating social conflicts.

Such arguments apply not only to humans, but also to human organisations. By constituting intermediate bodies, such non-human agents can better defend their cause, define industry norms and share good practices to self improve. Specialisation thus helps similar non-human agents build more resilient networks.

In software development too, intermediate bodies have emerged and have been essential to define industry norms,
and to increase the interoperability of various public and/or private software solutions.
Some prominent examples include the \emph{World Wide Web Consortium (W3C)}, 
the \textit{Internet Engineering Task Force (IETF)}
or the \emph{Web Hypertext Application Technology Working Group (WHATWG)}, 
which have shaped Internet and Web protocol standards,
thereby facilitating and securing the work of all companies 
that develop, exploit and depend on the Internet.

\subsubsection{The (positive) social chain reaction of specialisation.}

By its psychological and moral dimensions \cite{hughes1951mistakes}, specialisation involves social interactions. Thus, specialised tasks arising from division of labour are part of extended and complex ensembles involving professionals, as well as non-professionals \cite{hughes1956social}. Organized or not, they develop and defend (opposing) visions of what work should be. In this way, specialisation becomes part of social interactions during which legitimacy, monopolies, competition, and everything that delimits professional territories, are discussed. Specialisation is not fixed. It depends on interactions between stakeholders, and its limitations fluctuate. In the context of general AI, particularly generative AI promising to replace a considerable amount of professions and create new ones, adopting this vision of specialisation and division of labour which is not only technical, but also social, moral and psychological, would make it possible to anticipate and support emerging specialities.

\subsubsection{A simplified lesson from company towns.}

In an era where AGI is increasingly becoming the proclaimed goal of AI development, and when the most powerful billionaire are talking about the ``everything app''\cite{ArsTechnica}, it might be worth considering the 19th and 20th century use cases of company towns~\cite{garner1992company}. As defined in Wikipedia, ``a company town is a place where all or most of the stores and housing in the town are owned by the same company that is also the main employer. Company towns are often planned with a suite of amenities such as stores, houses of worship, schools, markets, and recreation facilities''. As a non-specialised, generalist, non-human entity; or as a general intelligence to stay in the parlance of AI, company towns can be seen as dealing with needs which are not limited to the sphere of work (in this case, postal, school, health, food, etc. services). Linked to what is known as ``industrial paternalism"~\cite{noiriel1988patronage}, this particular type of work management is associated with organisational problems. For example, the inability to provide all the services promised to workers, given that in fact, this system designed for blue-collar workers, mainly offered its advantages to white-collar workers. Another example is the difficulty of supervising workers, because it is impossible to be efficient by playing the roles of employer, priest, carer, policeman, etc., all at the same time, and therefore to bypass the division of labour and specialisation. Over and above the untenable nature of this strict social control of workers for reasons of labour profitability, this model of work organisation, based on task and power concentration, has not lasted. The reasons for the collapse of this system arguably include the replacement of the State by the management of the company town, the replacement of national law by the company practices, and most importantly, the replacement of the centuries-long process of specialisation of tasks and activities, by the one-company-do-it-all mindset of company-towns.

\section{A Case for Specification}
\label{sec:specification}

Closely related to specialisation,
\emph{specification} consists of precisely defining what a module excels at
(and what it cannot perform safely).
We dedicate a section to specification, 
as the value of specialisation is strongly tied to that of specification.

\subsection{Documentations}

The most straightforward step towards specification is \emph{documentation}.
To guarantee that any module or organisation is doing what it is supposed to be doing,
and more importantly that it is used as it is meant to be used,
it is essential to provide documents that describe the key features of the module or organisation.
Most products sold in developed countries must in fact be accompanied with such a documentation.

In practice, documentation may be in conflict with the quest for ready-to-play products.
In particular, viral adoption is often dependent on the ease with 
which the products can be used without relying on documentation.
What fraction of the population has read ChatGPT's documentation?
The flip side of this ease of use is unfortunately that such products are more likely to be misused,
e.g. to be deployed in applications where they are ill-suited.

In the context of machine learning,
increased calls for documentation have arisen, 
especially in the case of machine learning models through model cards~\cite{mitchell2019model} 
and in the case of datasets through data sheets~\cite{gebru2021datasheets}.

\subsection{Type Systems and Proofs of Correctness}

However, careful specification can yield much more significant improvements,
especially with regards to the resilience and security of complex systems.
In particular, modern programming languages provide 
sophisticated type systems~\cite{cardelli1996type,matsakis2014rust,gaher2024refinedrust},
which allow to directly encode the specifications of a module,
and mathematically prove that the module correctly implements the specifications
at compilation time.
Crucially, this allows to catch bugs \emph{before} deployment.

Moreover, type systems prevent misuses of a module,
for instance by specifying the nature of the inputs that the module allows.
Better yet, rich type systems can determine the guarantees that a module can provide,
given some properties of the inputs that are fed to it.

Unfortunately, learning systems do not lend themselves well to such specifications,
which hinders a thorough bug-catching procedure ahead of deployment.
Arguably, this might be why bugs in learning systems have often been rebranded as \emph{hallucinations}
in the context of generative algorithms.
But fundamentally, their ubiquity could be traced to the lack of specifications and of means 
to construct solutions that (provably) verify the specifications.

Now, it is noteworthy that type systems allow verification at compilation time, given the source code.
In practice, an additional challenge is to verify the correctness of a program,
given its compiled binary code.
To meet this challenge, the fascinating domain of \emph{verifiable computing}~\cite{ahmad2018primitives}.
In addition to multi-party computations~\cite{garg2022two},
where carefully specified operations and communications enable more secure information processing,
verifiable computing leverages powerful cryptographic primitives
like \emph{Succinct Non-Interactive Arguments of Knowledge} (SNARK)~\cite{thaler2022proofs},
which enable a powerful computer to prove the soundness of its output to a weak verifier.

Recently, there has in fact been much progress towards verifiable machine-learning computing,
at inference time~\cite{fan2024validcnn} (after deployment) 
and at training time~\cite{fan2024vericnn} (before deployment).
In particular, these solutions allow to envisage 
the effortless enforcement of some existing laws (like EU's AI Act),
e.g. by demanding that all commercialized AI systems prove 
that they were trained on legally obtained data 
(i.e. excluding copyrighted, sensitive or error-prone data).
In fact, remarkably, the proofs may be constructed in \emph{zero-knowledge},
i.e. without disclosing non-legally-binding proprietary data.

Nevertheless, verifiable computing for machine learning will remain necessarily restricted
to the characteristics of a learning systems that can be humanly \emph{specified}.
While this includes important aspects (e.g. training or inference integrity),
this inevitably excludes other considerations 
(e.g. ``correct" hate speech moderation).

\subsection{Hypertelia and the Pitfalls of Automated Task Specification}   

In biology, hypertelia \cite{wenn1874154} designates an exaggerated growth of certain organs in relation to their function, to the point of making them annoying for the animal and its entire species (cf. canines of the saber-toothed tiger, too heavy antlers of the deer or disproportionate tail of the peacock). Taken up in philosophy by Simondon \cite{simondon1958mode,simondon1980mode} for his analysis of technical objects, the notion expresses in this context the idea of an exaggerated specialisation, and a functional over-adaptation of technical objects. Adapting an object too much to a particular purpose and context can result in its inability to function well. Tools adjusted to very specialized circumstances can lose autonomy outside of their specific technical environment. 
Specialisation must be guided by proven mechanisms and not be an end in itself. If guided by human choices, these can be inspired by the specialisation that has taken place in the professions. The challenge of good specification will arise more intensely when the granularity of specialisations is finer, which is enabled by automation. This is where a risk of hypertelia can occur, without being easily relieved. The challenge here is to determine who or what meta-algorithm defines the speciality of each sub-algorithm.

\subsection{Two Limits of Specifications}
\label{sec:solomonoff}

Unfortunately, many critical information processing tasks 
(e.g. content recommendation, email drafting, language translation)
are extremely hard to specify.
We highlight more generally two reasons why not all systems can be fully specified.

First, some tasks are too complex to specify, 
in the sense of the Kolomogorov-Solomonoff complexity~\cite{solomonoff1960preliminary,kolmogorov1963tables}.
Formally, a formalized specification is a program that, given any program,
returns whether the program verifies the specification.
The Kolomogorov-Solomonoff complexity is defined as the shortest program 
that implements this formalized specification.
Unfortunately, it is conjectured that many tasks require extremely complex specifications,
in the sense that they cannot be formally described in less than a million lines of codes.
As an example, EU's AI Act is 144 pages long.
Yet, it is clearly far from formalized and has been argued to be very incomplete~\cite{laux2024trustworthy}.
Another example is today's web standards, 
which correspond to very long documents that specify languages like HTML, CSS and ECMAScript, among others.

Second, some specifications are extremely hard to agree on.
This is typically the case of content moderation,
but it also holds for text autocompletion, biasless image generation or content recommendation.
Given additionally the lack of knowledge 
about the human population's distribution of specification preferences,
and perhaps even about one's own preferred specification,
a choice of specification may seem premature and inappropriate.
This is one heart of the challenges surrounding ``AI alignment"~\cite{hoang2019fabuleux, el2024goodhart, majka2025strong}.

In the absence of concise and clearly consensual specifications,
should systems still aim to be specified?

\subsection{Specifying Unspecifiable Tasks Through Specified Governance}

We stress that the specification challenge is an old problem.
Throughout centuries, the correct punishment of a convicted felon has been extremely hard to specify.
Nevertheless, this does not mean that we should give up on the specification effort and, e.g.,
abandon the choice of punishment to an all-powerful judge or dictator.

Instead of specifying the correct punishment, 
our societies have worked hard on specifying \emph{how} to specify the correct punishment.
This may be called the problem of the specification of the \emph{governance} of specifications.
Democracies have typically solved it by writing constitutions.
which specified how any modification of the law could be enforced.
Similar governance structures are arguably needed for organizations and algorithms
whose goals are partially open-ended (and thus under-specified).

Remarkably, a growing line of research has provided new solutions for \emph{algorithmic governance}.
In particular, WeBuildAI~\cite{lee2019webuildai} proposed a software 
which allow a number of stakeholders to collaboratively select
the recipients of food donations.
This software was itself subject to both informal and formal specifications,
e.g. both donors, recipients, volunteers and the organising nonprofit association
should all have some voting power on the decision,
or they may either use the software's learning-based system to construct a model of their preferences,
or write themselves a model of their preferences.
This system has subsequently inspired other algorithmic governance systems,
e.g. for trolley dilemmas~\cite{noothigattu2018voting},
kidney donation~\cite{freedman2020adapting}
content recommendation~\cite{hoang2021tournesol},
proposal prioritization~\cite{small2021polis}
and contextual note selection~\cite{righes2023community}. More recently, \cite{solidago} aims to clarify the challenges of the specification
of the algorithmic collaborative governance of the scoring of any set of alternatives.
The list of subtasks to better specify includes 
participant verification,
trust propagation through a web of trust,
preference generalization,
and secure preference aggregation,
among others.

\section{Conclusion}
\label{sec:conclusion}

In this paper, we showed the limits of the arguments against specialisation, when applied to non-human information-processing entities.
We highlighted the industrial, democratic and security values of specialisation,
especially when it is accompanied with careful specification.
In particular, we articulated how the state of knowledge in adversarial machine learning,
complex system engineering, economics, and the sociology of work, occupations and organisations all point to 
the numerous issues of generality.
Finally, after acknowledging the limits of specifications in general,
we emphasised on the importance of specifying the governance of under-specified tasks,
especially when these tasks are complex 
or when they do not lend themselves to consensual specifications across populations and time.

We hope that the improved understanding of the value of specialisation will help
researchers, developers, managers, organisations, regulators and politicians
better orient the construction of a more prosperous,
more secure and more sovereign information space.

\section*{Acknowledgement}

The authors thank Peva Blanchard for fruitful comments.

\bibliography{references}

\end{document}